\documentclass[fleqn,usenatbib]{mnras}

\usepackage{newtxtext,newtxmath}

\usepackage[T1]{fontenc}
\usepackage{ae,aecompl}

\usepackage{graphicx}	
\usepackage{amsmath}	
\usepackage{amssymb}	

\newcommand{\be}{\begin{equation}}
\newcommand{\ee}{\end{equation}}

\title[Sodium density structure using laser guide stars]{Measuring line-of-sight sodium density structure using laser guide stars}

\author[J. A. Hellemeier et al.]{Joschua A. Hellemeier,$^{1,2}$\thanks{E-mail: jhelleme@phas.ubc.ca}
Domenico Bonaccini Calia$^{2}$,
Paul Hickson$^{1}$,
Angel Otarola$^{3}$
\newauthor and Thomas Pfrommer$^{2}$
\\
$^{1}$University of British Columbia, Department of Physics and Astronomy, 6224 Agricultural Road, Vancouver, B.C., V6T 1Z1, Canada\\
$^{2}$European Southern Observatory, Karl-Schwarzschild-Strasse 2, 8046 Garching bei Muenchen, Germany\\
$^{3}$TMT International Observatory, 100 West Walnut Street, Suite 300, Pasadena, CA 91124, USA
}
\date{Accepted XXX. Received YYY; in original form ZZZ}

\pubyear{2019}

\begin{document}
\label{firstpage}
\pagerange{\pageref{firstpage}--\pageref{lastpage}}
\maketitle

\begin{abstract}
The performance of adaptive optics systems employing sodium laser guide stars can be improved by continuously monitoring the vertical density structure of mesospheric sodium along the line of sight. We demonstrate that sodium density profiles can be retrieved by amplitude modulation of continuous wave (CW) lasers. In an experiment conducted at the Large Zenith Telescope (LZT), ESO's Wendelstein Raman-fibre laser was amplitude-modulated with a pseudo-random binary sequence and profiles were obtained by cross-correlation of the modulation pattern with the observed return signal from the laser guide star.  For comparison, high-resolution profiles were obtained simultaneously using the lidar system of the LZT. The profiles obtained by the two techniques show noise contamination, but were found to agree to within the measurement error. As a further check, a comparison was also made between several lidar profiles and those obtained by simultaneous observations using a remote telescope to image the laser plume from the side.
The modulated CW lidar technique could be implemented by diverting a small fraction of the returned laser light to a photon counting detector. Theoretical analysis and numerical simulations indicate that, for 50 per cent modulation strength, the sodium centroid altitude could be retrieved every 5 s from a single laser guide star, with an accuracy which would induce a corresponding wavefront error of 50 nm for the ELT and less than 30 nm for the TMT and GMT. If multiple laser guide stars are employed, the required modulation amplitude will be smaller. 
\end{abstract}

\begin{keywords}
instrumentation: adaptive optics -- telescopes
\end{keywords}



\section{Introduction}
All next-generation extremely-large telescopes (ELTs) will employ adaptive optics (AO) to reach unmatched sensitivity and angular resolution \citep{beckers1993}. ELTs will facilitate next-level observations in many areas of astrophysics \citep{skidmore2015}. AO systems mitigate the wavefront phase distortion induced by turbulence in the Earth's atmosphere, allowing near-diffraction-limited resolution. The distorted wavefront propagating from a reference star is sensed by a wavefront sensor (WFS) and the information is used to control one or more deformable mirrors (DM), conjugated at one or more turbulence layer altitudes. Natural guide stars (NGS) have the advantage of providing information on the tilt and focus components of wavefront distrotion. However, stars of sufficient brightness are rare and do not allow complete sky-coverage \citep{wang2009}. Therefore, ELTs will use laser guide stars (LGS) in conjunction with NGS to achieve adequate wavefront correction. 

LGS can be created either by Rayleigh backscattering in the lower atmosphere or by fluorescent excitation of sodium in the upper mesosphere and lower thermosphere \citep{foy1985,jeys1991}. For extremely-large telescopes, sodium LGS are better suited, as their higher-altitude reduces errors arising from the cone effect \citep{max1997}. ELTs will employ multiple LGS and tomographic reconstruction techniques to map the distribution of turbulence in real time. In particular, lasers tuned to the Sodium D$_{2a}$-line are used to excite transitions in the hyperfine structure of neutral sodium in the Earth's mesosphere. Since the excited atoms tend to migrate to the ground state of the D$_{2b}$-line, re-pumping by tuning a fraction of the laser light to the D$_{2b}$-line enhances the return flux for CW laser or long-pulsed laser excitation \citep{telle2008,holzlohner2010}. The column of luminous sodium in the mesosphere appears star-like when viewed from directly behind the laser beam  propagation axis, but when viewed with an offset, the finite vertical extent of the sodium layer causes the LGS to appear elongated \citep{thomas2010}. The intensity distribution within the LGS image reflects the vertical structure of the sodium region.

Since the 1980's the sodium region has been extensively studied using lidar systems \citep{clemesha1980}. These studies revealed seasonal and diurnal variations and a dependence of the sodium column density on latitude \citep{fan2007, gumbel2007, hedin2011}. The statistics of the sodium layer, from the point of view of laser guide star adaptive optics operation, have been studied for latitudes similar to those of the ESO VLT Paranal Observatory \citep{moussaoui2010}.

Advection of meteor ablation trails across the lidar beam results in sharp fluctuations in the sodium centroid altitude, and sporadic sodium layers are sometimes observed \citep{clemesha1996}. With advances in the design of AO systems, interest in the sodium layer grew, and the need for higher-resolution measurements became apparent. Performance estimates for ELTs, based on an extrapolation of existing data to typical AO update frequencies, \citep{davis2006} indicated that temporal fluctuations in the sodium density structure could degrade extremely-large telescope performance. This motivated higher-resolution studies using the LZT lidar system \citep{hickson2007, pfrommer2009, pfrommer2010a, pfrommer2014}. These revealed oscillatory and turbulent structure, arising from gravity waves and Kelvin-Helmholtz instabilities, that result in significant variability even on sub-second timescales. 

Temporal variability in the vertical structure of the sodium region produces variations in the intensity structure within the LGS image. This results in fluctuations in the LGS centroid position which lead to errors in the sensing of the atmospheric turbulence \citep{herriot2006}. The primary effect is a focus error caused by fluctuations in the sodium centroid altitude. Secondary effects result from variations in the internal structure, which induce higher-order errors. Since the elongation increases with increasing offset, this effect is more detrimental for large-aperture telescopes \citep{thomas2011}. 

To improve performance, AO systems need to be able to distinguish variations in the  centroid altitude from focus changes introduced by the atmosphere. The latter require AO correction, whilst the former should be ignored. One way to over come this problem is to continuously monitor the sodium density profile generating the LGS. 

More in general, major observatories using LGS-AO will benefit by monitoring the sodium layer profiles and their evolution on timescales of tens of seconds, and are therefore planning lidar monitoring systems. A simple method to obtain a sodium layer profiling system, a lidar, is by modulation of a fraction of the emitted laser amplitude as described in \citet{butler2003,she2011} or as in this paper.

Results from a sodium-profiling experiment using an amplitude-modulated CW laser are presented and the feasibility of this method for the next generation extremely-large telescopes is discussed. 

\section{Theory}

The modulated CW technique \citep{butler2003} can be understood as follows: Consider a single layer of sodium, at line-of-sight distance $z$, having infinitesimal thickness $dz$. The return flux will display the same modulation pattern as the transmitted beam, but with a time delay $2z/c$. The response from an extended sodium region is thus proportional to the integral of the modulation function $g(t) \in [0,1]$, weighted by the sodium density profile. Specifically, the return flux $s(t)$, in units of photons s$^{-1}$, is given by
\be
  s(t) = s_0 \int_0^\infty g(t-2z/c)f(z) dz. \label{eq:s1}
\ee
Here $s_0$ is the return flux when there is no modulation and $f(z)dz$ is the fraction of the received flux that originates from fluorescing sodium atoms  within the line-of-sight distance $z$ to $z+dz$. This fraction is proportional to the atomic sodium density $\rho(z)$ weighted by the inverse square of the distance (commonly referred to as ``flux weighted''),
\be
  f(z) ~\propto~ \rho(z)z^{-2}.
\ee
In writing Eqn. (\ref{eq:s1}), we have assumed that back-scattered light from the troposphere is blocked, so that the entire signal is due to resonant scattering from mesospheric sodium atoms.

The lower limit of the integral in Eqn. (\ref{eq:s1}) can be extended to $-\infty$ by defining $f(z)$ to be zero for negative $z$. Making the substitution $\tau = 2z/c$ we can write this as a convolution,
\begin{align}
  s(t) & = \frac{cs_0}{2} \int_{-\infty}^\infty g(t-\tau)f(\tau) d\tau, \nonumber \\
  & = \frac{cs_0}{2} g\ast f. \label{eq:s2}
\end{align}

If the modulation $g$ has the property that its autocorrelation function approximates a Dirac delta function $\delta(t)$,
\be
  g\star g \equiv \int_{-\infty}^\infty g(\tau)g(t+\tau)d\tau \simeq \delta(t), \label{eq:ac}
\ee
then we can recover the sodium density profile by computing the cross correlation between the modulation function and the return flux. Using Eqn. (\ref{eq:ac}) and well-known properties of the convolution and correlation, we find
\begin{align}
  g\star s & = \frac{cs_0}{2} g\star (g\ast f), \nonumber \\
  & = \frac{cs_0}{2} (g\star g) \ast f, \nonumber \\
  & \simeq \frac{cs_0}{2} \delta \ast f  = \frac{cs_0}{2} f. \label{eq:estimated profile}
\end{align}

\subsection{Recovering the sodium density profile}

In practice, we measure a series of discrete samples separated by the sampling time interval $\Delta t$. The number of photons detected in the interval centred at time $t_i$, will be denoted by $s_i $, where $i = 0, 1, 2, _{\cdots}$. Assuming that $\Delta t$ is much less than the timescale for temporal variations of the sodium density, $s_i = s(t_i)\Delta t$. 

In discrete form, Eqn. (\ref{eq:s1}) becomes
\be
  s_i = s_0 \Delta t \sum_{j = 0}^\infty g_{i-j}f_j. \label{eq:s2}
\ee
where $g_k = g(t_k)$ and $f_j = f(z_j)\Delta z$ and $z_j = j\Delta z$. Here $\Delta z = c\Delta {t}/2$ is the distance interval that corresponds to an interval $\Delta t$ in the round-trip light travel time.

We employ a modulation function that has the form
\be
  g_k = 1+\frac{\epsilon}{2}[p_k-1], \label{eq:g}
\ee
where $p_k$ is a pseudo-random binary sequence (PRBS) that takes the discrete values -1 or 1, and $\epsilon \in [0,1]$ is the modulation fraction. A modulation fraction $\epsilon$ will cause a decrease in laser power of $\epsilon / 2$. PRBS are pseudo-stochastic sequences generated by deterministic algorithms \citep{topuzouglu2006}. They are similar to random sequences, but do have some distinguishing properties. For a detailed description of PRBS, see \citet{sarwate1980} and references within. We assume that the PRBS is a maximum-length sequence, having $m = 2^l-1$ values, for $l$ being a positive integer. Such a sequence has the property that its circular autocorrelation is a delta function with a small constant offset,
\be
  r_{jk} \equiv \frac{1}{m}\sum_{i=0}^{m-1} p_{(i+j) \% m}p_{(i+k) \% m} = \frac{m+1}{m}\delta_{jk}-\frac{1}{m}, \label{eq:p_auto}
\ee
where $\delta_{jk}$, the Kronecker delta, takes the value 1 if $j = k$ and 0 otherwise, and the notation $n \% m$ denotes modulo arithmetic, $n$ mod $m$. We see from Eqn. (8) that the PRBS circular autocorrelation takes the value 1 when $j = k$ and $-1/m$ otherwise.

\begin{figure}
\includegraphics[width=\columnwidth]{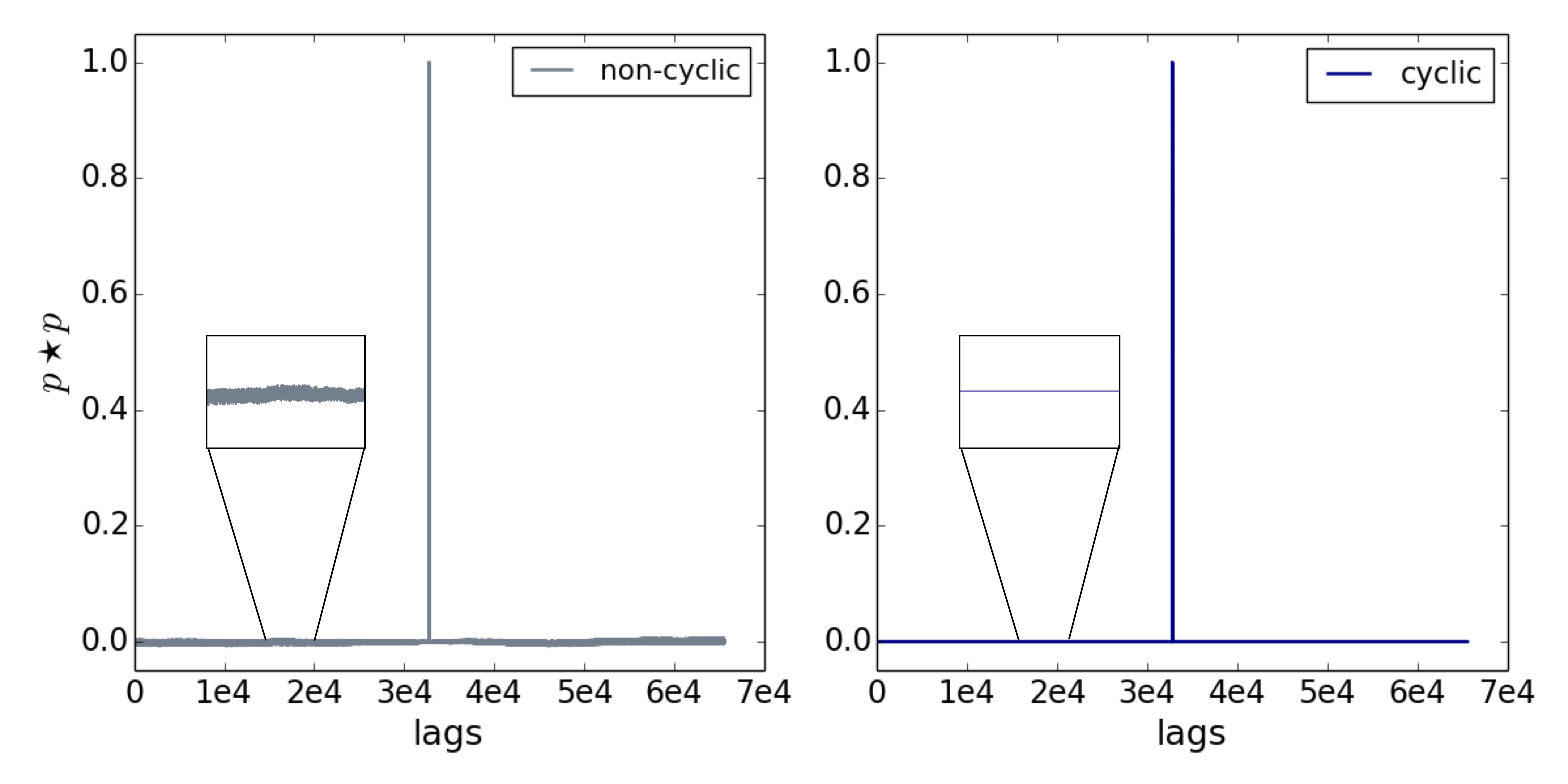}
\caption{Non-cyclic and cyclic autocorrelation  for a PRBS of length $2^{16}-1$. In the non-cyclic case there is a noise floor for the delta-like autocorrelation. In case of a cyclic autocorrelation, the noise vanishes and the autocorrelation is constant outside the spike. The value of the constant is the negative inverse of the length of the PRBS, in this case $-1/(2^{16}-1)$.    }
\label{fig:comp_lid_plume}
\end{figure}

The sequence contains $2^{l-1}$ positive and $2^{l-1} - 1$ negative unit values, so its mean value is
\be
  \bar{p} \equiv \frac{1}{m}\sum_{i=0}^{m-1} p_i = \frac{1}{m}. \label{eq:p_mean}
\ee

If the PRBS is repeated continuously, with no delay between successive copies, the circular correlation is equivalent to the standard correlation. We shall assume that this is the case, and shall therefore drop the modulo notation.

Let $z_0$ be the distance to the bottom of the sodium region and divide the propagation path through this region into $n$ intervals of size $\Delta z$. Let $l$ be the index that corresponds to the start of the sodium region, i.e. $l\Delta z = z_0$.
From Eqns. (\ref{eq:s2}) and (\ref{eq:g}),
\begin{align}
  s_i & = s_0\, \Delta t \sum_{j=l}^{n+l-1}  \left[1+\frac{\epsilon}{2}(p_{i-j}-1)\right]f_j. \label{eq:s3}
\end{align}

At any time $t_i$, we can make an estimate $\hat{f}_j$ of the sodium distribution by computing the cross correlation between the PRBS and the LGS signal. Consider the quantity
\be
    \hat{f}_j(t_i) = \frac{2}{\epsilon s_0 \Delta t (m+1)}  \sum_{k=0}^{m-1}(p_{i-k}-a) s_{i-k+j},
    \label{eq:density} 
\ee
where $m$ is a positive integer and $a$ is a constant. Replacing $j$ by $q$ in Eqn. (\ref{eq:s3}), substituting it into Eqn. (\ref{eq:density}) and using Eqns. (\ref{eq:p_auto}) and (\ref{eq:p_mean}), we obtain
\begin{align}
    \hat{f}_j(t_i) & = \frac{2}{\epsilon (m+1)}  \sum_{k=0}^{m-1} \sum_{q=l}^{l+n-1} (p_{i-k}-a) \left[1+\frac{\epsilon}{2}(p_{i-k+j-q}-1)\right]f_q, \nonumber \\
& =  \frac{m}{m+1}\sum_{q=l}^{l+n-1} \left[r_{jq} -a\bar{p} +  \frac{2-\epsilon}{\epsilon}(\bar{p}-a) \right]f_q, \nonumber \\
& =  \sum_{q=l}^{l+n-1} \left[\delta_{jq} - \frac{1+a}{m+1} +  \frac{(2-\epsilon)(1-am)}{\epsilon(m+1)}\right]f_q. \label{eq:hatf}
\end{align}
If we now set
\be
 a = \frac{2(1-\epsilon)}{\epsilon + m(2 -\epsilon)} \label{eq:a}
\ee
Eqn. (\ref{eq:hatf}) becomes
\be
 \hat{f}_j = f_j \label{eq:hatf2}
\ee
and the density profile is recovered from the return signal. 

\subsection{Performance}

An important question is the accuracy with which the density profile can be recovered. This depends on the integration time and LGS flux available. The primary source of noise arises from photon Poisson statistics. Accordingly, the best estimate of the variance in the number $s_i$ of photons detected in the $i$-th bin is $\text{var} \, s_i = s_i$. 

The variance in the estimated profile obtained from a single PRBS sequence can be found from Eqn. (\ref{eq:density}) and (\ref{eq:a}), 
\begin{align}
    \text{var}\,\hat{f}_j & = \frac{4}{[\epsilon s_0 \Delta t (m+1)]^2}  \sum_{k=0}^{m-1}(p_{i-k} -a )^2\text{var}\, s_{i-k+j}, \nonumber \\
    & = \frac{4m}{[\epsilon s_0 \Delta t(m+1)]^2} \left(1- \frac{2a}{m} + ma^2 \right)\bar{s}\Delta t, \nonumber \\
    & \simeq \frac{4 \bar{s}}{(\epsilon s_0 )^2 T }, \label{eq:varhatfphoton}
\end{align}
where $\bar{s}$ is the average flux and $T = m\Delta t$ is the integration time. In obtaining the last equation we have assumed that $m \gg 1$ and have neglected terms of order $m^{-2}$.

From Eqn. (\ref{eq:s3}), 
\be
  \bar{s} \simeq s_0 \frac{2-\epsilon}{2}.
\ee
Therefore, 
\be
\text{var}\,\hat{f}_j  \simeq  \left(\frac{2-\epsilon}{\epsilon}\right)^2 \frac{1 }{ N_\gamma}. \label{eq:varhatfphoton2}
\ee
where $N_\gamma = \bar{s}T$ is the total number of photons detected. Thus the expected RMS error in $f_j$, due to photon noise, is
\be
  \sigma_f = \frac{2-\epsilon}{\epsilon} N_\gamma^{-1/2},
\ee
We see that this is independent of $j$. The noise is the same for all bins, and does not depend on the density profile. It depends only on the number of photons detected, and the modulation amplitude. From this it is clear that if the signal is integrated over multiple repeats of the PBRS, the same result applies if $N_\gamma$ is taken to be the total number of photons received during the integration time.

Let us now consider a parameter of central importance for AO, the line-of-sight distance to the centroid of the sodium distribution,
\be
  \bar{z} = \int_0^\infty f(z) z dz. \label{eq:barz}
\ee
Variations in this distance result in wavefront sensing errors, primarily in the focus term. The resulting RMS wavefront error, averaged over the telescope pupil, is (\citet{davis2006}, \citet{herriot2006})
\be
  \sigma_{wfe} = \frac{1}{16\sqrt{3}} \frac{D^2}{\bar{z}^2}\sigma_{\bar{z}}. \label{eq:wfe}
\ee
where $D$ is the telescope aperture diameter. An important goal of real-time sodium density profiling is to estimate $\bar{z}$ so that AO wavefront sensing algorithms can be updated to accommodate rapid temporal variations in the sodium distribution. 

We wish to estimate the accuracy with which we can measure $\bar{z}$. In discrete form, the direct estimate of $\bar z$ is 
\be
  \hat{z} = \frac{1}{w} \sum_{j = l}^{l+n-1} z_j \hat{f}_j,
\ee
where
\be
  w = \sum_{j = l}^{l+n-1} \hat{f}_j.
\ee
The variance of this estimate is
\begin{align}
  \text{var}\, \hat{z} & = \sum_{j=l}^{l+n-1} \left|\frac{\partial \hat{z}}{\partial \hat{f}_j}\right|^2 \text{var}\, \hat{f}_j, \nonumber \\
  & = \frac{1}{w^2} \sum_{j=l}^{l+n-1} \left(z_j-\hat{z}\right)^2 \text{var}\, \hat{f}_j,  \nonumber \\
  & = \sum_{j=l}^{l+n-1} \left(z_j-\hat{z}\right)^2 \text{var}\, \hat{f}_j.\label{eq:varhatz} 
\end{align}
The last step follows because the expected value of $w$ is unity (from the definition of $f$).

Substituting Eqn. (\ref{eq:varhatfphoton2}) into Eqn. (\ref{eq:varhatz}), we obtain
\begin{align}
    \text{var}\,\hat{z} & \simeq  \left(\frac{2-\epsilon}{\epsilon}\right)^2 \frac{1 }{ N_\gamma} \sum_{j=l}^{l+n-1} \left(z_j-\hat{z}\right)^2, \nonumber \\
  & \simeq  \left(\frac{2-\epsilon}{\epsilon}\right)^2 \frac{nZ^2 }{12 N_\gamma}. \label{eq:var_z}
\end{align}
Taking the square root, we obtain the estimated error in the centroid,
\be
   \sigma_z  \simeq  \frac{2-\epsilon}{\epsilon}\sqrt{\frac{n }{12 N_\gamma}}Z. \label{eq:sigma_z}
\ee
To see how this depends on the modulation amplitude, observe that from Eqn. (\ref{eq:s3}), the number of detected photons $N_\gamma$  is proportional to $1-\epsilon/2$. 
Therefore, 
\be
  \sigma_z ~\propto~ \frac{1}{\epsilon}\sqrt{1-\epsilon/2}.  \label{eq:sigma_z II}
\ee
  
We see from Eqn. (\ref{eq:var_z}) that the variance in the centroid estimate is proportional to $n$. It can be reduced by sampling more coarsely. However, one must also consider the error resulting from the discrete sampling. If the density profile contains spatial frequencies higher than the Nyquist frequency $f_N = 1/2\Delta z$, aliasing will occur. The sampling rate should be chosen to be high enough to prevent aliasing. This will place a lower limit on the resolution and accuracy that can be achieved. 

Both the flux and the extent $Z$ of the sodium region depend on zenith angle $\zeta$, as does $\bar{z}$. If the sampling rate is kept constant, $n$ will be proportional to the extent Z. In terms of the airmass, $X \simeq \sec\zeta$, we find
\begin{align}
  Z & ~\propto~ X, \\
  n & ~\propto~ X, \\
  \bar{z} & ~\propto~ X, \\
  N_\gamma & ~\propto~ X\exp(-0.921 k X),
\end{align}
where $k$ is the extinction coefficient at a wavelength of 589 nm and 0.921 is more accurately $2.5/\ln(10)$. From Eqns. (\ref{eq:wfe}) and (\ref{eq:var_z}), we expect that the wavefront error will vary with airmass according to
\be
  \sigma ~\propto~ X^{-1}\exp[0.460 k (X-1)].
\ee
This function decreases with increasing airmass. Therefore, the greatest error will occur for observations at the zenith. 

\section{Simulations}

\begin{figure}
\includegraphics[width=\columnwidth]{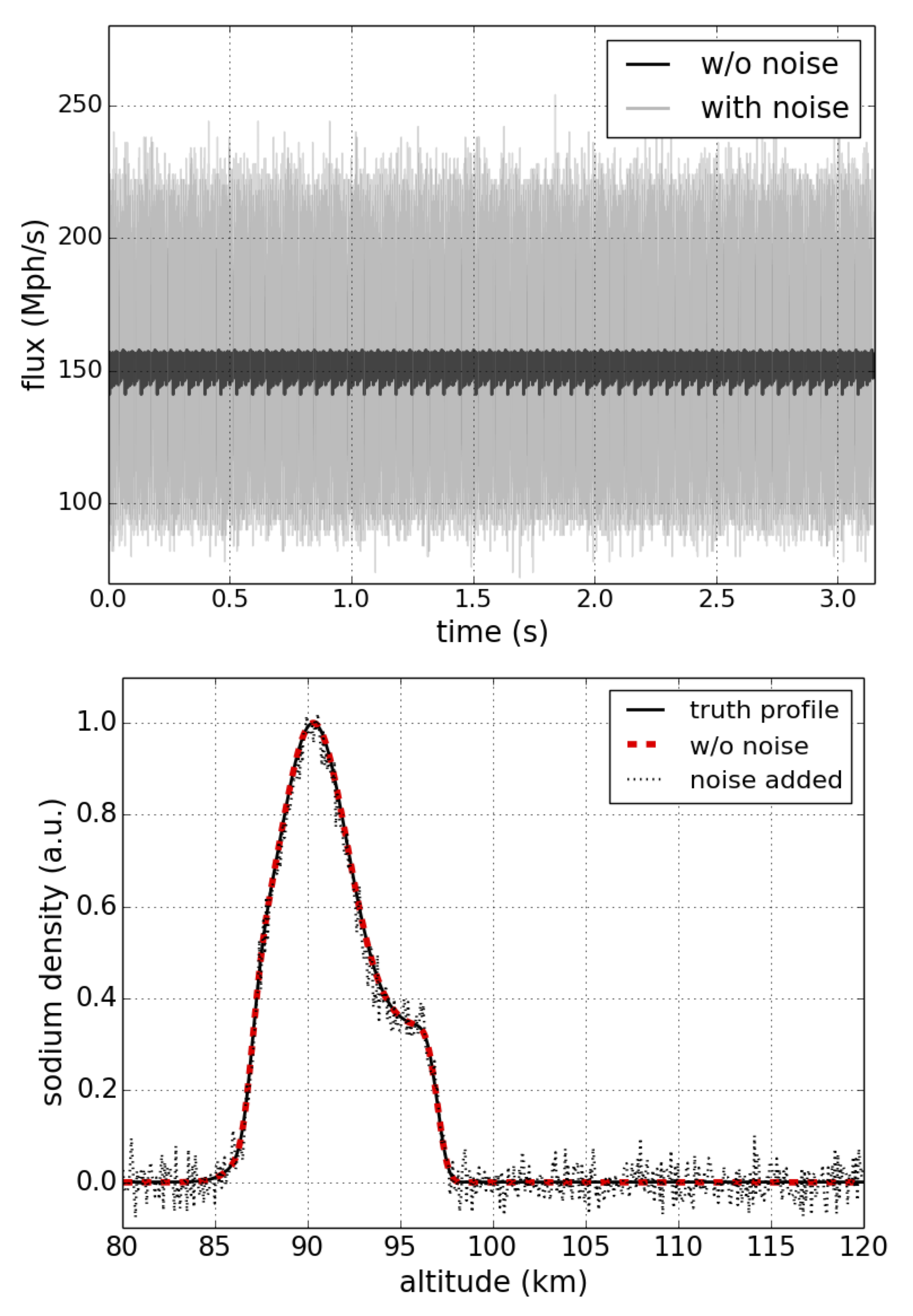}
\caption{ \textit{Top:} Simulated flux without and with added noise for a repeated PRBS of length $2^{17}-1$ and a sampling rate of 2 MHz. \textit{Bottom:} Comparison of the retrieved profiles, the profile obtained from a signal without added noise (red dashed line) matches the truth profile well. The dotted line shows the result when Poisson noise is added.  }
\label{fig:rf-prof}
\end{figure}

In order to validate the theoretical predictions and to assess the feasibility of the method for the ELT, numerical simulations were performed. A smooth model sodium profile was generated which was similar to real sodium profiles observed in the experiment at the Large Zenith Telescope. The model profile\footnote{The results of the simulations were tested for different profile shapes. The results were consistent for different profile shapes. The discussion presented here only includes one profile shape.} served as the true profile and is shown in Fig. \ref{fig:rf-prof}. From the convolution of the true profile with the laser signal (Eqn. \ref{eq:s2}) the return signal was calculated. The amplitude of the laser power was partially-modulated by a PRBS and Poisson noise was added to the return signal. The estimated sodium profile was then obtained from the cross-correlation of the  return flux and the PRBS (Eqn. \ref{eq:estimated profile}). 

\begin{table}
	\centering
	\caption{Paramters used for wavefront error calculation.}
	\label{tab:parameters_wfe}
	\begin{tabular}{lc} 
		\hline
		parameter & value \\
		\hline
		LGS magnitude & 7.0 \\
		flux at primary mirror      & 15.8 Mph/m$^{2}$/s\\
		telescope throughput      & 0.3  \\
		fraction of leakage light  & 0.03 \\
		diameter ELT     & 39.3 m  \\
		diameter TMT     & 30.0 m  \\
		diameter GMT    & 25.4 m \\
		\hline
	\end{tabular}
\end{table}

The parameters employed for the simulations are listed in Table \ref{tab:parameters_wfe}. The assumption of 3 per cent of the light available to retrieve the sodium profile corresponds to typical fraction of light that leaks past the LGS beam splitter in the AO optics. Fig. \ref{fig:rf-prof} shows an example of the simulated return signal and retrieved density profile for the ELT. The simulations were performed with a range of PRBS length, modulation strength, sampling rate and total integration time. As the total integration times exceed the period of one modulation sequence of the laser power by a single PRBS, the laser power was repeatedly-modulated by the same PRBS till the desired integration time was reached.

\begin{figure}
\includegraphics[width=\columnwidth]{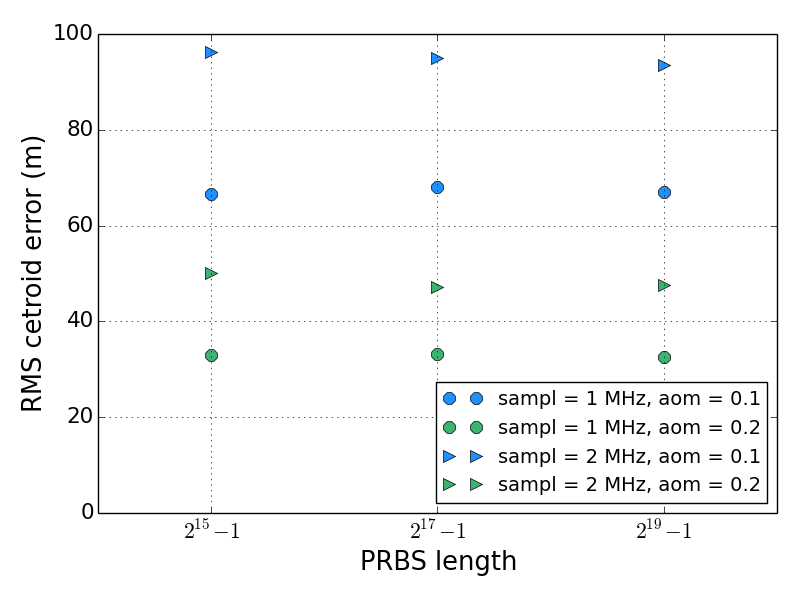}
\caption{Simulated RMS centroid error for the ELT, for different sampling rates, vs. PRBS length. The RMS centroid error does not depend on PRBS length.}
\label{fig:RMS-cen vs prbs}
\end{figure}

\begin{figure}
\includegraphics[width=\columnwidth]{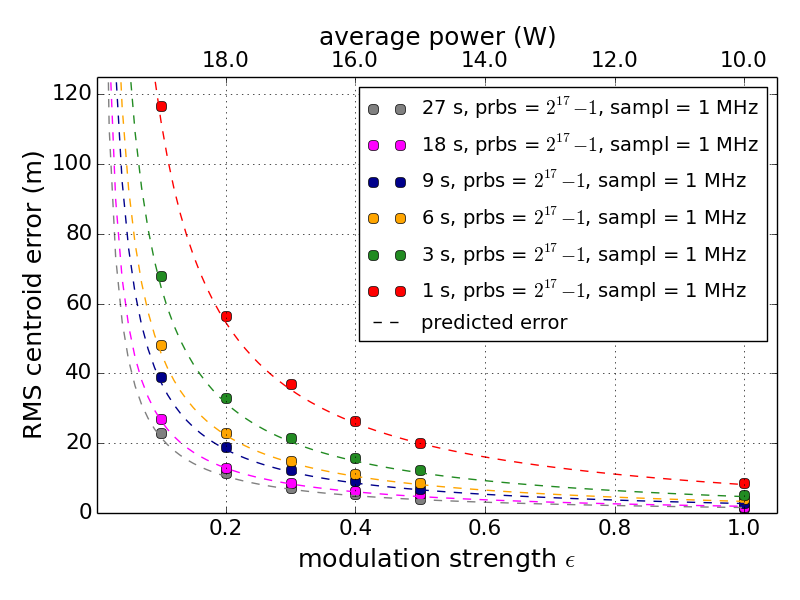}
\caption{ Comparison of the analytically-predicted (dashed lines) and simulated (dots) RMS centroid error for profiles retrieved by the CW lidar method on the ELT. The predicted and simulated centroid error agree well.}
\label{fig:RMS-cen vs aom}
\end{figure}

For each set of parameters (modulation depth, sampling rate, PRBS length), 1000 estimated retrieved sodium density profile for the same truth profile were simulated for a telescope aperture equal to that of the ELT. From these estimated density profiles the RMS error of the centroid was determined. The results for the simulated RMS centroid error were compared to the theoretically-predicted centroid errors (Eqn. \ref{eq:sigma_z}). The results are shown in Figs. \ref{fig:RMS-cen vs prbs} \& \ref{fig:RMS-cen vs aom} and confirm the results of the theoretical analysis. As predicted, the RMS centroid error of the simulations does not depend on PRBS length. The theoretical-predicted strong dependence of the RMS centroid error for low modulations strengths has been verified with the simulations, see Fig. \ref{fig:RMS-cen vs aom}. For low modulation strength, an increase in modulation strength yields the highest increase in centroid accuracy and the overall dependency of the RMS centroid error on modulation strength follows a behaviour proportional to $(1/\epsilon)\sqrt{1-\epsilon/2}$, as predicted by Eqn. (\ref{eq:sigma_z II}). 

\section{Feasibility for extremely-large telescopes}
The knowledge of the vertical profile of the sodium density at the mesosphere is
useful as support to the observations in general, and may furthermore allow avoiding the use of a natural guide star to sense the atmospheric induced focus error.

Indeed if the mesospheric sodium vertical profile centroid can be known a priori, the atmospheric focus contribution to the wavefront error could be measured directly from the LGS wavefront sensor, without the need of a natural guide star. In this case, the sodium profiles centroid has to be obtained with sufficient accuracy, within a fraction of the rate of change of the sodium profile.

\begin{figure}
\includegraphics[width=\columnwidth]{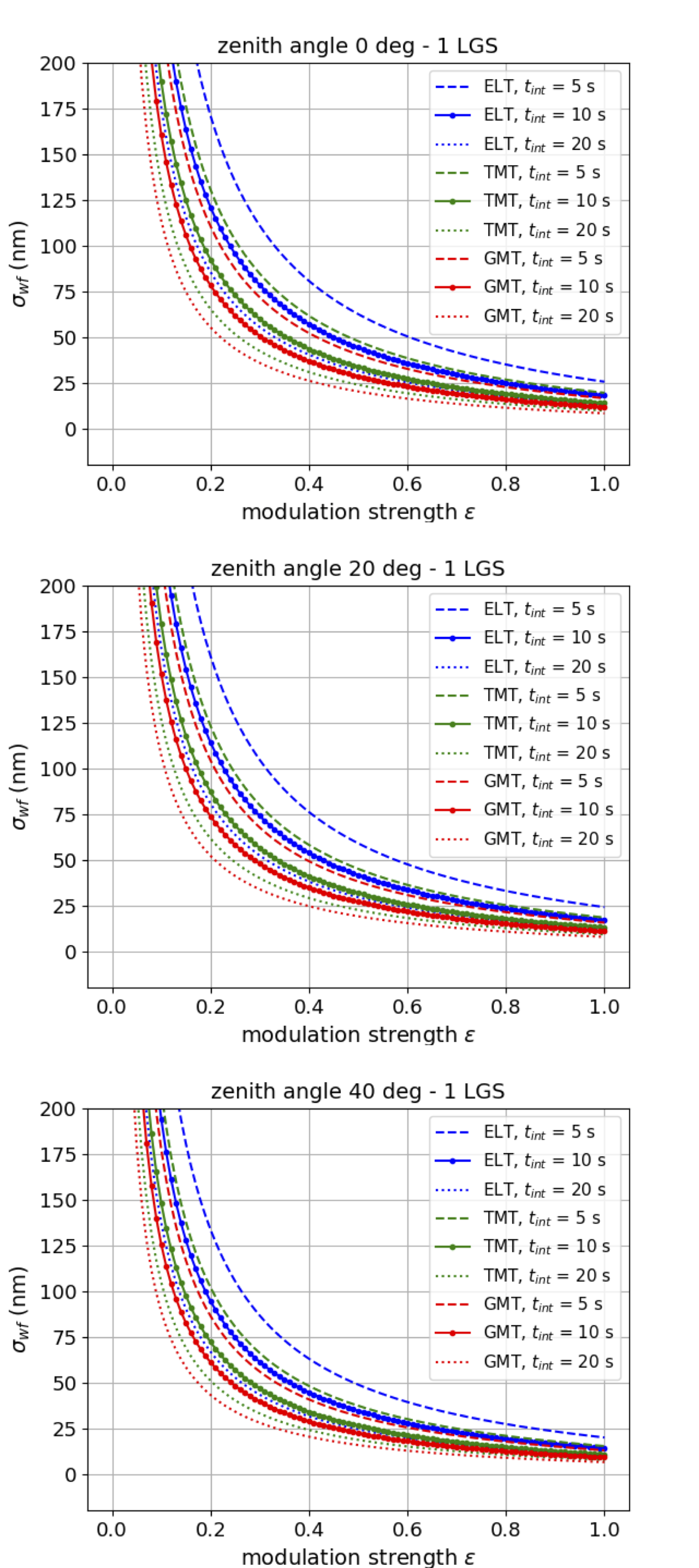}
\caption{Induced wavefront error from inaccuracy in sodium centroid estimation for extremely large telescopes at different zenith angles (\textit{top:} 0$^{\circ}$ zenith angle, \textit{middle:} 20$^{\circ}$ zenith angle, \textit{bottom:} 40$^{\circ}$ zenith angle).}
\label{fig:1-LGS}
\end{figure}

\begin{figure}
\includegraphics[width=1.\columnwidth]{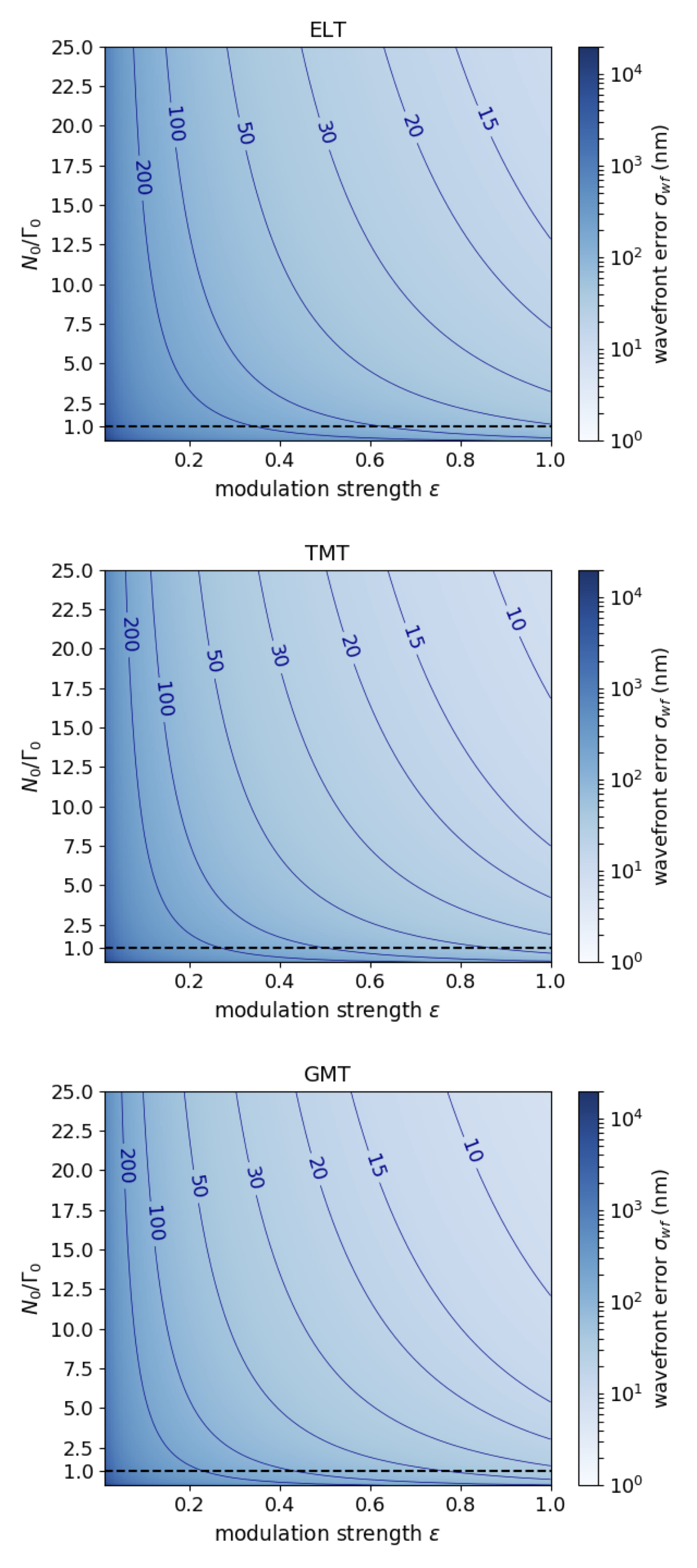}
\caption{ Contour plots of the RMS focus error induced by errors in the estimate of the sodium centroid altitude, for 20$^{\circ}$ zenith angle. The normalization factor $\Gamma_{0}$ was determined for a flux at the primary mirror of $1.58\times10^7$ photons m$^{-2}$ s$^{-1}$, corresponding to a LGS brightness of 7th magnitude, a telescope throughput of 0.3, a fraction of leakage light of 3 per cent and an integration time of 1 second. The dashed horizontal line indicates where $N_{0}$ is equal to $\Gamma_{0}$.}
\label{fig:wfe-2d}
\end{figure}

RMS wavefront focus term errors can be determined from the sodium profile centroid errors obtained by the simulations. Substituting Eqn. (\ref{eq:sigma_z}) into Eqn. (\ref{eq:wfe}) gives
\be
  \sigma_{wfe} = \frac{1}{16\sqrt{3}} \frac{D^2}{\bar{z}^2} \frac{2-\epsilon}{\epsilon}\sqrt{\frac{n}{12N_{\gamma}}} Z. \label{eq:wfeII}
\ee
From this, the functional dependance of the RMS wavefront error on modulation strength, aperture diameter, sampling rate and total number of photon was estimated. Results for the ELT, the TMT and the GMT are presented in Figs. \ref{fig:1-LGS} and \ref{fig:wfe-2d}.  In Fig. \ref{fig:1-LGS}, the RMS wavefront focus error is shown for a single LGS and a range of integration times, modulation strengths and zenith angles. The corresponding focus wavefront error decreases with increasing zenith angle. It can be seen that for an integration time of $t_{int} = 5$ s corresponding wavefront errors below 50 nm can be reached by modulation of a single LGS by 0.6 on the ELT and by 0.4 on the TMT and GMT. A modulation of 0.6 and 0.4 would decrease LGS return flux by 30 per cent and 20 per cent respectively. From Eq. (\ref{eq:wfeII}) it is obvious that the corresponding wavefront error depends on the total number of photons $N_{\gamma}$ and the modulation strength $\epsilon$. However, the parameter $N_{\gamma}$ itself depends on the modulation strength. We introduce a parameter $N_{0}$ which is independent of modulation strength. $N_{0}$ is the number of photons at 0 per cent modulation strength. It is related to the total number of photons by $N_{\gamma} = N_{0}(1-\epsilon/2)$. $N_{0}$ is dependent on integration time, LGS return flux, telescope throughput, fraction of leakage light used and number of LGS used. For MCAO or MOAO systems one could consider using multiple laser guide stars for retrieving the sodium profile \footnote{Unpublished results for the horizontal structure function of sodium density profiles show small variations in the profiles for small angular separations. So multiple LGS in MCAO or MOAO systems could be used.}. 
    
The estimated wavefront error, for the three extremely-large telescopes for observation at a zenith angle of 20$^{\circ}$, is shown in Fig. \ref{fig:wfe-2d}. In the figure the quantity $\Gamma_{0}$ is used as a normalization factor and is given by the number of photons at 0 modulation strength for an integration time of $t_{int} = 1$ s, a LGS return flux at the primary of $1.58\times10^7$ photons m$^{-2}$ s$^{-1}$, a telescope throughput of 0.3, a fraction of leakage light of 0.03 and one LGS being modulated. For low values of $N_{0}$ ( < $2.5 \Gamma_{0}$) the contours run horizontally. An increase in $N_{0}$ will result in smaller estimated wavefront errors. For large values of $N_{0}$ ( > $10 \Gamma_{0}$), the contours run almost vertically. An increase in $N_{0}$ will not result in significantly smaller estimated wavefront errors. To reduce the estimated wavefront error the modulation strength needs to be increased. The case of a wavefront error below 50 nm, as discussed earlier, could not only be achieved by one LGS and modulation strength of $0.4 - 0.6$, but also using six LGS and modulation strength of $0.2 - 0.3$. For the future when higher LGS return flux becomes available, scenarios which yield wavefront errors below 20 nm are possible. For example, for six LGS with doubled return flux and an integration time of two seconds, yielding a ratio of $N_{0}/\Gamma_{0} = 24$, for a modulation strength of 0.6 the wavefront error decreases below 20 nm for each of the extremely-large telescopes. 


\section{Experimental set-up and observations}

\begin{figure}
\includegraphics[width=\columnwidth]{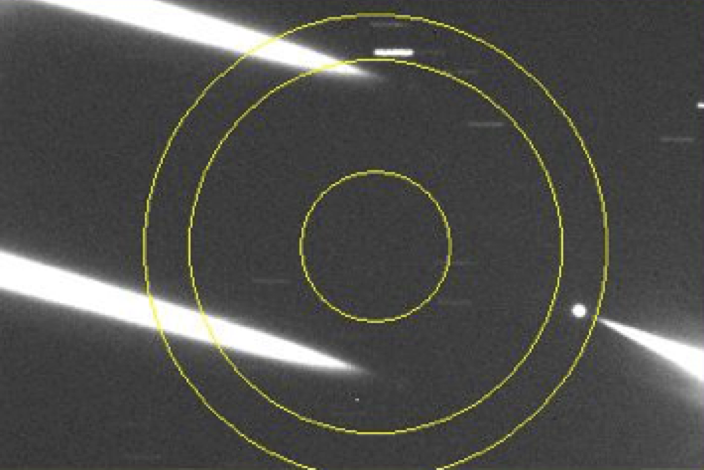}
\caption{Image of the LGS of the lidar and the CW laser. The CW LGS is in the lower right corner. Two beams are visible from the lidar laser, which was chopped in pointing at a rate of 50 Hz, in order to check for horizontal variations in the sodium structure. The outer yellow ring indicates the boundary of the sensitive area of the avalanche photodiode. 
}
\label{fig:2lasers}
\end{figure}

\begin{figure}
\includegraphics[width=\columnwidth]{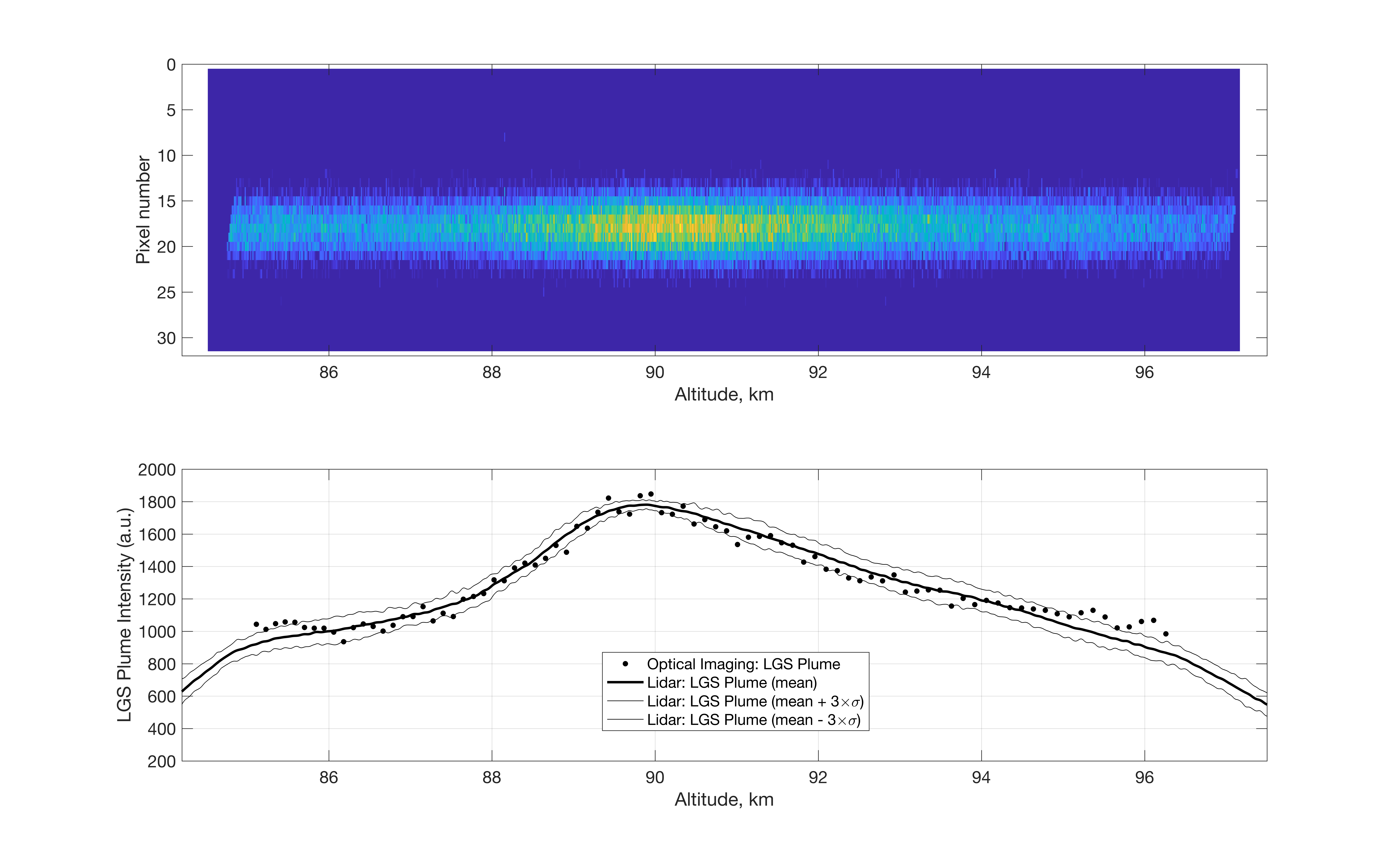}
\caption{LGS plume image obtained by sideview observation (top) and a comparison of the lidar density profile with the profile obtained by the sideview observation (bottom). These profiles were obtained during the preparation phase of the experiment. Both profiles shown are flux-weighted profiles, as an AO system would see them.}
\label{fig:comp_lid_plume}
\end{figure}

The experiment was conducted at the LZT, located near Vancouver, Canada, in July 2014. Preliminary results have been published in \citet{hellemeier2018}. The ESO Wendelstein laser \citep{calia2012} and the pulsed laser of the LZT lidar system were used to simultaneously measure sodium density profiles. Those of the LZT lidar serve as a reference by which the accuracy of the CW method can be assessed. The LZT lidar system employed a Nd:YAG pumped dye laser, having a power of $4-5$ W, and a pulse length of  6 ns \citep{pfrommer2010b}. The ESO Wendelstein laser is based on a Raman fiber amplifier, providing a maximum output power of $20$ W CW at 589nm. 

An acousto-optic modulator was used to apply the PRBS modulation. Both lasers were tuned to the sodium $D_{2a}$ line. 10\% of the laser power of the CW laser was used to repump the sodium $D_{2b}$ line.  All observations were conducted at the zenith. 

The backscattered light from both lasers was collected by the LZT and imaged onto a large-area avalanche photodiode (APD) detector operating in analogue mode.  A narrowband interference filter having a central wavelength of 600 nm and a half-power bandwidth of 25 nm was inserted in front of the APD to reduce background light. The APD signal was amplified by a Femto DHCPA current amplifier having a transimpedance gain of 1 V/$\mu A$ and a bandwidth of 1.8 MHz and was then digitized at a rate of 16.6 MHz. 

Fig. \ref{fig:2lasers} shows an image of the zenith sky taken using a 30 cm auxiliary telescope and a CCD camera. The CW LGS was placed near the edge of the sensitive area of the APD in order to prevent detector saturation from the bright Rayleigh-scattered light from the CW transmitted beam. The lidar beam was chopped at 50 Hz between two positions on the sky, in order to simultaneously measure the horizontal structure function of the sodium centroid altitude.  The results of that program will be the subject of a separate publication. It will suffice to remark that no significant variations were seen during the observations reported here. 

The amplitude modulation of the CW laser was synchronized with the 50 Hz lidar pulses. The two lasers were projected alternately in order to avoid overlap of the signals. The output of the CW laser was enabled 3 ms after the lidar pulse, after the backscatter from the sodium region was received. The PRBS was then applied for 10 ms, after which the CW laser was disabled until the next cycle. Two amplitude modulation depths were employed, corresponding to 67.5 per cent and 50 per cent of the maximum power. Two PRBS with different sampling rates were used for the test. 

The detector, amplifier electronics and cables were carefully shielded and grounded to reduce radio-frequency interference. A two-pole low-pass filter having a 1-MHz 3-db cutoff was inserted at the input of the digitizer. However, the signal still showed residual noise, mostly in the range from 20 to 100 kHz. 

Observations were conducted in the night of the 17th July, 2014, as outlined in Table \ref{tab:observation}. On an earlier night the profiles of the lidar were compared to the sideview profiles obtained with a 20 cm telescope located 2.4 km south of the LZT. This test was done to provide an independent validation of the lidar profile. The lidar and the sideview profiles show good agreement with differences being less than three standard deviations (Figure \ref{fig:comp_lid_plume}). 

\begin{table}
\centering
\caption{Observation details }
\begin{tabular}{lcccc}
\hline
date & start time & end time & AOM & PRBS \\
(UTC) & (UTC) & (UTC) & & sampling rate \\
\hline
17 Jul	& 06:26 & 07:58 & 67.5\% & 1 MHz \\
17 Jul	& 09:14 & 09:48 & 50.0\% & 1 MHz \\
\hline
\end{tabular}
\label{tab:observation}
\end{table}

\section{Data analysis}

The lidar profiles were obtained by inverting the amplified APD signal and then subtracting the background. This was done by first fitting a sixth-order polynomial to the signal in the altitude range from 75 to 140 km, which during the season of observation extends beyond the boundaries of the sodium region. The background at each altitude in this range was then estimated by linear interpolation between the end points of the polynomial function. This was then subtracted to provide the final profile. Line-of-sight distances were converted to altitude above mean sea level by adding the LZT site altitude of 400 m. 

\begin{figure}
\includegraphics[width=\columnwidth]{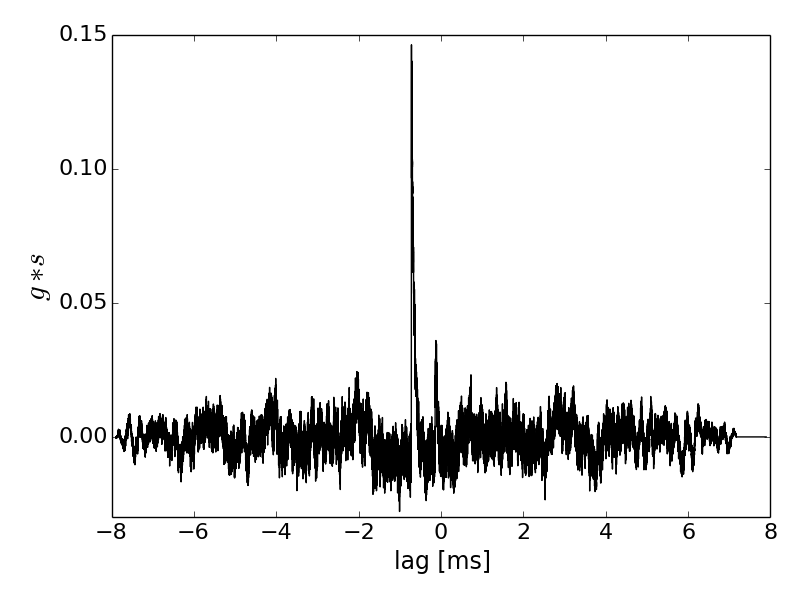}
\caption{Example of the cross-correlation of the CW signal with the PRBS. The distinct peak originates from Rayleigh-backscattering, the smaller peak on the right of it from the sodium fluorescence.}
\label{fig:crosscor}
\end{figure}

A similar technique for background subtraction was used for the CW profiles. A seventh order polynomial was fit to the signal in the $75 - 140$ km altitude range, and the background was modeled by a linear function matching the values of the polynomial at the endpoints of the altitude range. The use of polynomials of higher order was necessary to fit low-frequency noise components in the background. In order to additionally reduce noise, the signal from the CW laser was smoothed with a Savitzky-Golay filter \citep{savitzky1964}, before the cross-convolution with the PRBS sequence was applied. The filtering window contained 50 data points, which corresponds to a period of $\sim$ 3 $\mu$s. A polynomial function of third order was used as the fitting function for the Savitzky-Golay filter.  

An example of the cross-correlatied signal is shown in Figure \ref{fig:crosscor}. Two  distinct peaks are clearly seen. The larger peak originates from Rayleigh scattering, while the smaller peak is the sodium resonance signal. 

\begin{figure}
\includegraphics[width=\columnwidth]{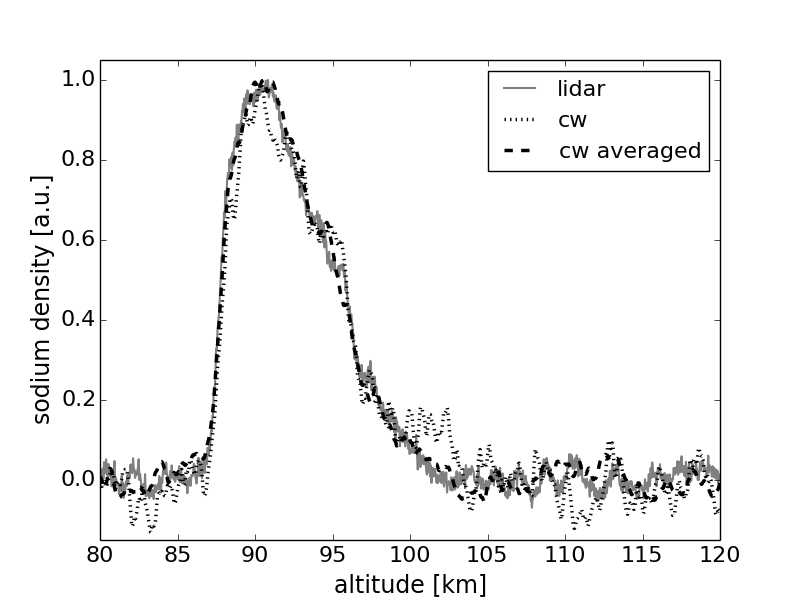}
\caption{Comparison of a lidar (solid) and CW (dotted) profile. The amplitude modulation was 0.67. The dashed line shows the CW profile obtained from averaging over a period of 10 seconds.} 
\label{fig:comp_prof}
\end{figure}

The final lidar and CW profiles are the median of 127 consecutive laser shots, and 127 transmissions of the PRBS sequence, respectively. This is equivalent to a continuous data acquisition (integration time) of 1 second. Both profiles were scaled to have unit area. An example of lidar and CW profiles is shown in Figure \ref{fig:comp_prof}.

There was some jitter in the time delay between the lidar pulse and the start of the CW laser modulation, which resulted in a small random altitude shift for the CW profile. In addition, the power spectra of our signal showed radio-frequency noise which was probably picked up from nearby radio stations. The random shift in altitude and the radio-frequency noise are experimental flaws which were caused by our experimental setup. The shift in altitude was different for every PRBS transmission and could be up to 6 $\mu$s (900 m). For each profile, the shift was estimated by minimizing the second moment of the difference between the CW and the lidar profiles. In order to reduce noise, a mean CW profile was obtained by averaging profiles within $\pm 5$ s. This period is expected to be long enough to cancel low-frequency noise components, but short enough that sodium structure variations should be small.

The CW profiles were compared to the lidar profiles by testing the similarity of four key measures. In addition to the estimated centroid altitude $\hat{z}$, \citet{pfrommer2014} specify the square root of the second moment $\upsilon$, the width of enclosed energy $w$ and the width of a specified flux threshold $\xi$ as useful metrics. The three measures characterize shape variations in the sodium profile. The width of enclosed energy is the width containing a fixed percentage of the integrated flux. The width for the flux threshold is the altitude range over which the profile exceeds a fixed percentage of the maximum flux. Due to the noise contribution, conservative thresholds of 80 per cent for the enclosed energy and 20 per cent for the flux were chosen. For all comparisons the flux-weighted profiles and not the density-weighted profiles were used.  

\section{Results}
\begin{figure}
\includegraphics[width=\columnwidth]{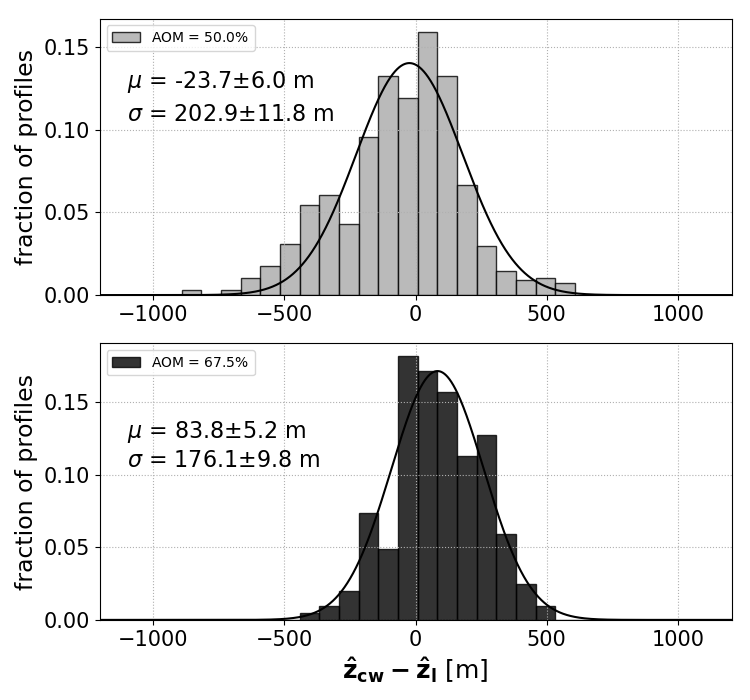}
\caption{Distribution of the differences of the centroid altitudes of the CW and lidar profiles.}
\label{fig:centroid}
\end{figure}

\begin{figure}
\includegraphics[width=\columnwidth]{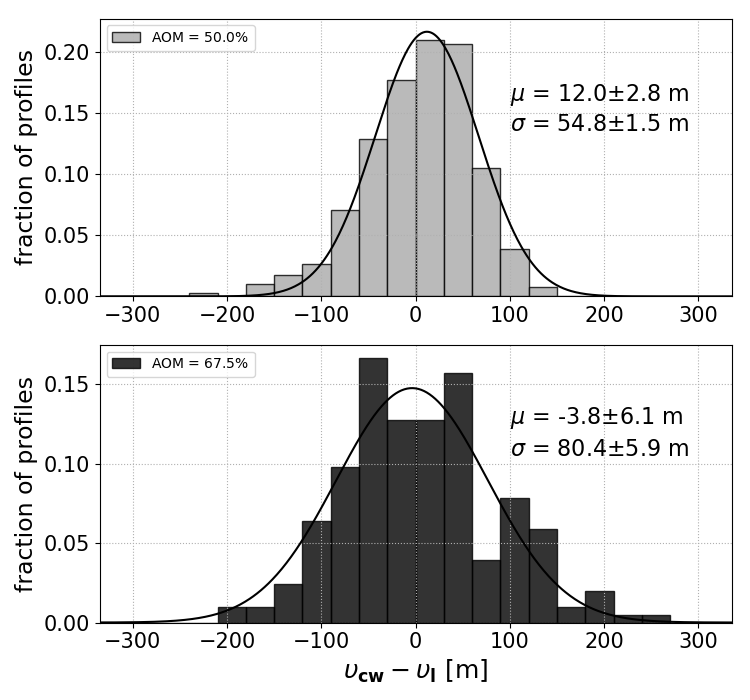}
\caption{Distribution of the differences in the square root of the second moment of the profiles.}
\label{fig:mom2}
\end{figure}

\begin{figure}
\includegraphics[width=\columnwidth]{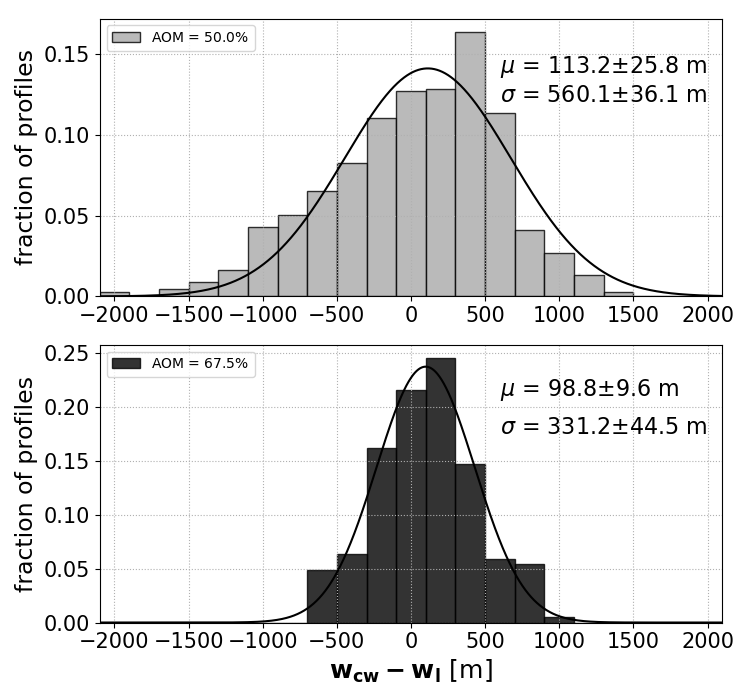}
\caption{Distribution of the differences of the widths of 80\% included energy.}
\label{fig:w80}
\end{figure}

\begin{figure}
\includegraphics[width=\columnwidth]{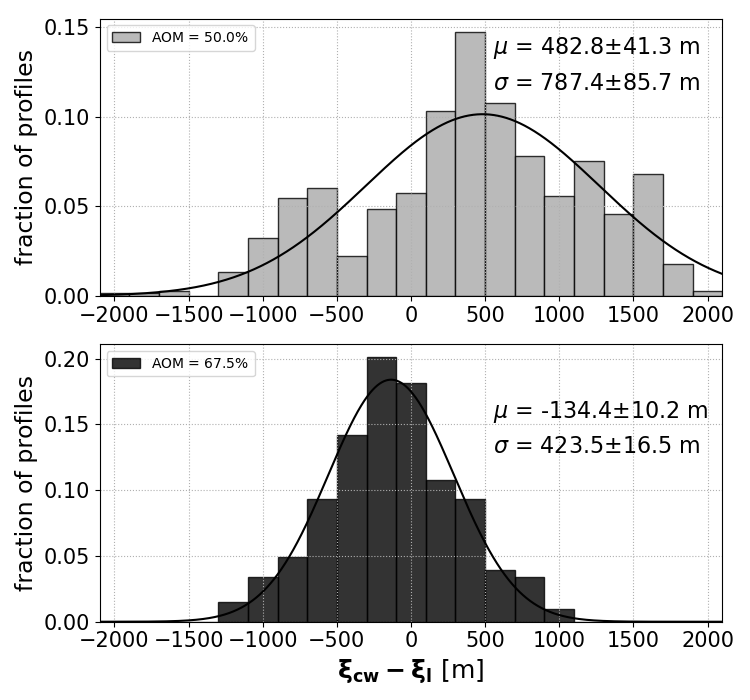}
\caption{Distribution of the differences of the widths exceeding 20\% of the maximum flux.}
\label{fig:f80}
\end{figure}

A total of 882 ten-second profiles were obtained from the CW laser: 678 profiles used 50 per cent amplitude modulation and 204 used 67.5 per cent amplitude modulation depths. Each profile was compared with a one-second averaged lidar profile acquired at the same time. The offset $\Omega$ was calculated by subtracting the 'truth' measure $x_{l}$ from the measure of the CW profile $x_{cw}$,
\begin{equation}
    \Omega =x_{cw}-x_{l} \, . 
	\label{eq:offset}
\end{equation}
Histograms of the offsets between the parameters for the lidar and CW profiles are shown in Figs. \ref{fig:centroid} - \ref{fig:f80}. In order to reduce the impact of  outlying points, gaussian curves were fit, by the least-squares method, to the distributions. The mean $\mu$ and standard deviation $\sigma$ of the gaussian curves are indicated in the figures. A summary of the fit parameters can be found in Table \ref{tab:fit_param}.

For the distribution of differences in the centroids, the standard deviation $\sigma$ is 202.9 m for 50 per cent modulation and 176.1 m for 67.5 per cent modulation. From Eq. (\ref{eq:sigma_z}), we would have expected a RMS difference in centroid altitude for the experiment at the LZT of $20-40$ m for the modulation depths used. The significant larger differences in centroid altitude most likely result from afore-mentioned experimental effects . The widths $w$ and $\xi$ are typically about 10 km. The standard deviation of the differences in the 2nd moments is smaller than that of deviation for the enclosed energy width $w$ and the width for 20 percent flux threshold $\xi$ by nearly an order of magnitude. It is likely that the widths $w$ and $\xi$ are more strongly affected by the signal robust noise, which has a greater impact in regions of the profile where the signal is weak.        

\begin{table}
\centering
\caption{\bf Parameters of the Gaussian fits.}
\begin{tabular}{lcrrrr}
\hline
parameter & $AOM$ & $\Omega_{\hat{z}}$& $\Omega_{\upsilon}$ & $\Omega_{w}$ & $\Omega_{\xi}$  \\
\hline
$\mu$ [m] & 50.0\% & -23.73 & 12.01 & 113.19 & 482.81 \\
$\mu$ [m] & 67.5\% & 83.76 & -3.75 & 98.78 & -134.43 \\
$\sigma$ [m] & 50.0\% & 202.88 & 54.75 & 560.13 & 787.36 \\
$\sigma$ [m] & 67.5\% & 176.09 & 80.44 & 331.24 & 432.50 \\
\hline
\end{tabular}
\label{tab:fit_param}
\end{table}

\section{Discussion \& Conclusions}

The sodium layer CW lidar is relatively simple to implement on telescopes, and could readily be used to monitor sodium profiles on timescales of tens of seconds, during operation. It could further be used for real-time sodium density profiling, with the ELT or other large telescope, by directing a small fraction of the light to the photon counter of the profiling subsystem. In an AO system, typically about 97 per cent of the light of the LGS is directed towards the WFS by a beam splitter. The remaining 3 per cent of the light from the LGS follows the path to the science camera. This fraction could be notch-filtered and redirected to a high-speed photon counting detector, used to measure sodium density profiles.

While the retrieved profiles taken at the LZT demonstrate that the CW lidar profiles are consistent with the 'standard' lidar profiles, our results obtained with a large area APD sensor used in linear mode were affected by a robust radio-frequency interference. In a photon-noise limited case the CW lidar method should have provided profiles with an accuracy of $20-40$ m RMS centroid offset. Instead, the spread found in our experimental data is on the order $150 - 200$ m. We believe that this discrepancy is due to electro-magnetic interference in the linear APD signal detected.

Our theoretical analysis and the numerical simulations show the potential of the method. The method could yield sodium centroid profiles with high accuracy. Different scenarios could be realized, either one LGS modulated with a higher modulation strength ($> 60$ per cent) or multiple LGS modulated with low modulation strength ($\approx 30-40$ per cent) could be used. We note that laser amplitude modulation strengths of $30-40$ per cent result in $15-20$ per cent decrease in the LGS return flux, which can be acceptable already with current 22W CW sodium lasers. 

Recent experiments done at the ESO WLGSU in La Palma \citep{haguenauer2019} have demonstrated the feasibility of splitting the uplink laser beam in two, to create two LGS angularly separated. The subsequent reduction of 50 per cent of the LGS flux corresponds to the loss of power due to a modulation of 100 per cent ($\epsilon = 1 $), in our CW lidar.

Calculations in the frame of the ESO Mavis project Phase A study indicate that using the 20W CW lasers in the VLT Galacsi LGS-AO system, with 20x20 cm sub-apertures Shack-Hartmann wavefront sensor, will still be able to support  MCAO systems at visible wavelengths.

Considering also that R\&D work is in place at ESO to both, enhance the LGS return flux by frequency chirping, and by increasing the laser power beyond 50 W at 589 nm, there is a good outlook for the CW lidar method to be used in operations, turned on when necessary. As this is the time of designing the extremely-large telescope AO systems, we solicit the system experts to take provisions for the CW lidar implementation into account.

Besides providing sodium laser profiles and abundances on timescales of several seconds, of general use during observations, by tuning the operational parameters of a CW lidar and assessing the consequences at LGS-AO system level, it can be conceived that the CW lidar provides the sodium layer profile centroid distance and the sodium profiles at sufficient speed,  to eliminate the need of NGS to sense focus, and to precisely work with matched filters, respectively. These aspects have to be analyzed more in detail, at system level and specifically for the different extremely-large telescope projects, in our opinion.

We see, for the parameters assumed in the ELT case, that for integration times on a sub-second scale it is not possible to retrieve estimated wavefront errors below 30 nm, unless the fraction of leakage light or a different parameter for the accessible flux is changed. We assume that the ELT error budget for the focus term is on the order of 30 nm. However, the information on the profile could still be updated several times per second, by writing the retrieved data into a continuously-updating buffer, which then can be evaluated for a set number of seconds of integration time. 

Matched-filter algorithms coping with the elongated LGS in Shack-Hartmann wavefront sensors could also benefit from real-time sodium density profiles obtained with integration times of several seconds. An analysis of the performance increase for wavefront sensing with matching-filter algorithms is an open task for the future.  

At present in AO systems, NGS are used for focus and tip-tilt sensing. If a method was developed to sense focus from the LGS, e.g. with the CW lidar sensing the centroid location in the sodium profile, or by side monitoring an extremely elongated LGS, the need for NGS to sense focus might be eliminated. LGS-AO systems which do not need NGS for regular operations will have 100\% sky coverage. Elimination of the need for NGS in LGS-AO operations is the strategy behind current ESO lead LGS-AO R\&D developments which aim at eliminating tip-tilt sensing needs from NGS as well, and the basis of our current development work.

\section*{acknowledgements}
\noindent This work was supported by grants from the Natural Sciences Engineering Research Council of Canada and the Canada Foundation for Innovation, and by the ESO funding of the Technology Development Program on LGS-AO R\&D. The TMT project provided also information and support.  We thank R. Holzl\"{o}hner for fruitful discussions and comments on PRBS and F. Rigaut, M. Thallon and J.-P. Veran for helpful discussions. 




\bibliographystyle{mnras}
\bibliography{biblo}


\bsp	
\label{lastpage}
\end{document}